\documentclass{aa}  

\usepackage{graphicx}
\usepackage{txfonts}
\usepackage{lipsum}
\usepackage{lscape}             % to rotate a single page table, example in appendix.
\usepackage{placeins}           % useful with \FloatBarrier, to keep 
\usepackage[colorlinks=true, linkcolor=blue, citecolor=blue, urlcolor=blue]{hyperref}
\def\PGPU{$\varphi-$GPU}

\newcommand{\be}{ \begin{equation} }
\newcommand{\ee}{\end{equation}}

\begin{document}

\title{Dynamical evolution of Milky Way globular clusters on the cosmological timescale} 
\subtitle{II. Terzan 2, 4, and 5 mass loss and collision tracking}
%   \subtitle{GC vs GC collisions}

   \author{M. Ishchenko\inst{1,2,3}
        \and 
            P. Berczik\inst{2,3,1}
        \and 
            D. Kuvatova\inst{3,4}
        }

   \institute{Main Astronomical Observatory, National Academy of Sciences of Ukraine, 27 Akademika Zabolotnoho St, 03143 Kyiv, Ukraine\\
             \email{\href{mailto:marina@mao.kiev.ua}{marina@mao.kiev.ua}}
        \and
            Nicolaus Copernicus Astronomical Centre, Polish Academy of Sciences, ul. Bartycka 18, 00-716 Warsaw, Poland
        \and 
            Fesenkov Astrophysical Institute, Observatory 23, 050020 Almaty, Kazakhstan
        \and 
%    Szechenyi Istvan University, Space Technology and Space Law Research Center, H-9026 Gyor, Egyetem ter 1. Hungary
%        \and 
            Farabi University, al-Farabi ave. 71, 050040 Almaty, Kazakhstan\\}

   \date{Received September 30, 20XX}

% \abstract{}{}{}{}{}
% 5 {} token are mandatory
 
  \abstract
  % context heading (optional)
  % {} leave it empty if necessary  
{The dynamical evolution of Galactic globular clusters is one of the key topics in modern astrophysics. It is also essential to understand their mutual gravitational interactions, especially in dense central regions, for reconstructing the assembly history of the Milky Way.}
  % aims heading (mandatory)
{We investigate the long-term dynamical evolution of Ter2, Ter4, and Ter5, focusing on their mutual interactions, mass-loss behaviour, and survivability in the dense Galactic centre environment. We expect that low-mass clusters are particularly more sensitive to gravitational perturbations from massive neighbours, and our study seeks to quantify these effects under realistic orbital and dynamical conditions.} 
  % methods heading (mandatory)
{We performed a suite of high-resolution direct $N$-body simulations over 8 Gyr, modelling three individual clusters that we also modelled as combined systems. Orbital reconstructions and dynamical analyses were carried out in a realistic time-evolving Galactic potential. We compared reference runs of isolated clusters with simulations of the full three-cluster system to quantify possible differences in mass loss, potential energy, and orbital behaviour. Pairwise collision statistics were extracted to assess the frequency and impact of possible close encounters.}
  % results heading (mandatory)
{Our simulations reveal multiple close encounters between the Terzan clusters. The most significant encounters occur between Ter2--Ter4 and Ter4--Ter5, with their tidal radii exceeding the minimum separation. A notable case is the pair Ter2--Ter4, which approaches within 10 pc at a relative velocity of $\sim$320 km/s. Overall, Ter4 is the most dynamically active cluster in close interactions. We found that the mass-loss rate is higher for the low-mass Ter2 and Ter4 systems in the combined three-cluster simulations than in our similar isolated runs, highlighting the importance of mutual cluster interactions. Differences in potential energy evolution further confirm that collective modelling alters the dynamical pathways of individual clusters. The common run clearly demonstrates that mutual gravitational interactions between clusters drive significant triaxial deformations, especially for Ter2 and Ter5, which evolve from nearly spherical to distinctly prolate shapes. In contrast, the isolated runs show clusters that remained almost perfectly spherical, confirming that the observed shape changes are correlated with the mutual interactions.}
{The survivability and dynamical evolution of Galactic centre globular clusters cannot be fully understood without accounting for collective interactions among all systems within a few kiloparsecs. Low-mass clusters are particularly sensitive to gravitational perturbations from massive neighbours, which accelerate their mass loss and alter their orbital histories. Our results emphasise the necessity of complex multi-cluster modelling in realistic Galactic potentials to capture the long-term fate of surviving and dissolved clusters.}


   \maketitle

\nolinenumbers
%%%%%%%%%%%%%%%%%%%%%%%%%%%%%%%%%%%%%%%%%%%%%%%%%%%%%%%%%%%%%%
\section{Introduction}{\label{sec:Intr}}
%%%%%%%%%%%%%%%%%%%%%%%%%%%%%%%%%%%%%%%%%%%%%%%%%%%%%%%%%%%%%%
The globular cluster (GC) system of the Milky Way (MW) includes more than 160 observable objects. One of the most complete and recent information about them was provided by the catalogue of \cite{Baumgardt2021} \footnote{\url{https://people.smp.uq.edu.au/HolgerBaumgardt/globular/parameter.html}}, which  contains the structural parameters for 168 objects identified as GCs. The masses of most GCs lie in the range of $\sim$$10^4$--$10^6$ M$_\odot$, and their half-mass radii (r$_{\rm hm}$) fall within 2--20 pc. Since the ages of GCs are comparable to the age of the Galaxy itself, they are definitely one of the best indicators in its dynamical evolution, including the formation of various structures and the whole assembly history \citep{Antonini2012, Kruijssen2020}. Some of the GCs have a pronounced heterogeneous stellar population, that is, multiple stellar populations \citep{Bica2016}. This might indicate not only the self-enrichment processes, but also the intense interactions with a surrounding environment.
One popular explanation for the presence of multiple stellar populations in Galactic GCs \citep[see][and references therein]{Bastian2018,Gratton2019} is the external origin of chemically distinct gas or stellar material, which may also result from close interactions caused by potential mergers of individual clusters \citep{Ferraro2009, Origlia2011, Khoperskov2018, MastrobuonoBattisti2019}. Direct GC close encounters or even collisions might seem extremely rare. However, according to several studies, such interactions might be fairly probable for GCs in the central MW region. For instance, \cite{Miocchi2006} modelled the evolution of massive and compact GCs in a triaxial gravitational potential and showed that although the impact on internal structure and orbital evolution from even face-on GC collisions is negligible, GC merging events into nuclear star cluster-like systems are potentially possible due to dynamical friction in the very inner regions of the Galaxy \citep{CapuzzoDolcetta2008a, CapuzzoDolcetta2008b, Dana2026}. According to \cite{Khoperskov2018} and \cite{MastrobuonoBattisti2019}, in a system of about 100 massive ($\sim$10$^7$ M$_\odot$) GCs, a few physical collisions can occur for each billion years, along with several close passages even with possible mass exchange. Their simulations included cases of GC mergers into a single system within 0.5 Gyr. In the long term, the resulting clusters lose mass and become more compact. When the merging clusters have similar dynamical properties, their stellar populations mix rapidly; otherwise, clusters with differing densities produce a system with multiple populations and distinct density profiles.

In this context, numerical studies of GC populations indicate that modelling clusters in a joint system leads to structural differences compared to their isolated evolution, primarily through tidal interactions and repeated close passages \citep{Khoperskov2018}. However, the strength of this effect is highly sensitive to the spatial configuration of the system and is expected to be most significant in dense environments or during the early evolutionary stages \citep{Miocchi2006}. \cite{Combes1999} further argued that clusters undergoing substantial mass loss can develop elongated shapes by precessing around the z-axis. While this does not constitute direct evidence that simultaneous modelling of multiple GCs necessarily leads to deviations from spherical symmetry, it suggests a possible link: if interactions within a GC system enhance the mass-loss efficiency, they might indirectly promote morphological transformations of this type.

In our previous studies \citep[hereafter, \hyperlink{I23}{\color{blue}{Paper~I}}, and \hyperlink{I232}{\color{blue}{Paper~III}}]{Ishchenko2023a, Ishchenko2023c}, the orbital modelling of GCs backward in time for 10 Gyr showed that some GCs complete a significant number of revolutions around the Galactic centre (e.g. $\sim$620 for Terzan 4 and $\sim$540 for NGC 6624). Several such clusters are associated with regions such as the central bulge and disk. The long-term presence of several objects within a relatively small Galactic centre volume can naturally lead to a certain number of mutual encounters. In our study in \hyperlink{I232}{\color{blue}{Paper~III}}, we numerically modelled 147 GCs in five selected time-variable MW-like potentials (TVPs) extracted from the cosmological database IllustrisTNG-100 \citep{Nelson2018} with the aim to conduct a statistical analysis of the likelihood of close encounters and possible collisions among GCs. It was found that most of these events occurred in the central region of the Galactic disk, within $\sim$4 kpc of the centre, with two ring-like zones with an increased encounter density observed at radii of about $\sim$1 and $\sim$2 kpc of the Galactic centre (see Fig. 8 in \hyperlink{I232}{\color{blue}{Paper~III}}). Across all modelled potentials, the rate of close passages ($\leq$50 pc) was estimated as $\sim$10 events every billion years. Moreover, actual collision events were detected for 22 GC pairs with a very high likelihood. Some of the most active participants in these events are Terzan 4, Terzan 2, NGC 6624, NGC 6440, Terzan 5, and Liller 1, all of which belong to the MW central GC subsystem and are located near the Galactic centre (see the detailed Table 3 in \hyperlink{I232}{\color{blue}{Paper~III}}). For a more detailed analysis of collisions and close passages, we here selected the Terzan cluster family: Terzan 2 and Terzan 4 as the most active collisional pair in all our five MW-like potentials. Terzan 5 was added to the list because of its high probability of close passages with Terzan 4 and also with Terzan 2. Terzan 5 is also one of the most massive clusters known today, with $\sim$10$^6$ M$_\odot$. These GCs were directly modelled as a full $N$-body system with up to several million particles each and with a detailed stellar evolution.

Terzan 2 is a relatively metal-rich GC ([Fe/H] = -0.86), while Terzan 4 is moderately metal poor ([Fe/H] = -1.38). Both belong to the inner bulge population \citep{Schiavon2024} and exhibit evidence of multiple stellar populations \citep{Kunder2021}. However, the most prominent example of a chemically complex stellar system is Terzan 5. This GC shows a wide spread in iron abundance ($\sim$1 dex) and a significant stellar age spread of approximately 5--8 Gyr. Its metallicity distribution is clearly multi-modal \citep{Massari2014}, with three distinct iron peaks: an extremely high-metallicity population (29\%, [Fe/H] $\approx$ 0.25), a moderately metal-rich population (62\%, [Fe/H] $\approx$ -0.3), and a relatively metal-poor population (6\%, [Fe/H] $\approx$ -0.8). This complexity indicates a highly inhomogeneous stellar population.

Several studies have questioned whether Terzan 5 is a genuine GC at all \citep{Origlia2011, Bailin2022}. Its orbit and chemical properties clearly suggest an in situ origin within the Galactic bulge, however \citep{Origlia2025}. The observed population complexity might partially be attributed to self-enrichment processes and the formation of subsequent stellar generations \citep{Ferraro2009, Valcarce2011}. However, the predicted mass fraction of second-generation stars is insufficient to explain the observed peculiar features. Alternative scenarios invoke a more complex dynamical history. For example, the presence of multiple populations could result from the merger of two originally distinct stellar systems \citep{Gavagnin2016, Khoperskov2018, MastrobuonoBattisti2019, Pfeffer2021}. 

Based on our previous findings in \hyperlink{I232}{\color{blue}{Paper~III}}, we carried out a more detailed dynamical analysis here, including a separate and common orbital-evolution modelling of the Terzan cluster family. We analysed the individual and collective mass loss of the clusters and as the energy evolution of their mutual gravitational interaction. 

In accordance with this, the paper is organised as follows. In Sect. \ref{sec:modell} we introduce our $N$-body reconstruction of the chosen GCs. In Sect. \ref{sec:mass-loss} we estimate the mass loss of the GCs during their evolution. Section \ref{sec:gc-timescale} analyses the GCs collision probability on cosmological timescales. In Sect. \ref{sec:tail} we present additional individual simulations to investigate the mutual gravitational effect of GCs. Finally, in Sect. \ref{sec:con} we summarise our results and draw the main conclusions.

%%%%%%%%%%%%%%%%%%%%%%%%%%%%%%%%%%%%%%%%%%%%%%%%%%%%%%%%%%%%%%%%%%%%%
\section{Full $N$-body reconstruction of Terzan 2, 4, and 5}\label{sec:modell}
%%%%%%%%%%%%%%%%%%%%%%%%%%%%%%%%%%%%%%%%%%%%%%%%%%%%%%%%%%%%%%%%%%%%

%%%%%%%%%%%%%%%%%%%%%%%%%%%%%%%%%%%%%%%%%%%%%%%%%%%%%%%%%%%%%%%%%%%%%
\subsection{Time-variable potential implementation}\label{subsec:tng}
%%%%%%%%%%%%%%%%%%%%%%%%%%%%%%%%%%%%%%%%%%%%%%%%%%%%%%%%%%%%%%%%%%%%

To enhance the physical realism of our simulations, we employed data from the cosmological magnetohydrodynamical  IllustrisTNG-100 simulation \citep{Nelson2018}. One of the key advantages of this cosmological simulation database is that the time-dependent gravitational potentials for a large number of different individual galaxies can be extracted.

From the entire set of simulated galaxies in TNG, we selected the MW-like object with ID {\tt 411321}, which at zero redshift ($z = 0$) exhibits properties that most closely match the observed parameters of today's MW. In particular, the selection was based on the similarity in total mass, as well as the spatial scales of the disk and halo components. The sampling and fitting procedures for this selected potential and for other MW-like potentials are discussed in detail in Table 1 and Fig. 2 of \hyperlink{I23}{\color{blue}{Paper~I}}. The code routines for sampling and fitting the selected potentials are also publicly available on GitHub\footnote{The ORIENT: \\~\url{ https://github.com/Mohammad-Mardini/The-ORIENT}}.

For the full $N$-body integration of the GCs, we used the popular parallel high-order code \PGPU\footnote{$N$-body code \PGPU: \\~\url{ https://github.com/berczik/phi-GPU-mole}}, which employs a fourth-order Hermite integration scheme with hierarchical individual block time-steps. This direct body code has proven itself well in modelling GCs in our previous papers \citep{Ishchenko2024, Weis2025, Dana2026}. An external potential {\tt 411321} was especially incorporated into the code. In addition, a stellar evolution library was implemented \citep{Banerjee2020, Kamlah2022a, Kamlah2022b}, enabling the treatment of the main stellar evolution processes during the dynamical modelling.

%%%%%%%%%%%%%%%%%%%%%%%%%%%%%%%%%%%%%%%%%%%%%%%%%%%%%%%%%%%%%%%%%%%%%
\subsection{Initial models for Terzan 2, 4, and 5}\label{subsec:ini-cond}
%%%%%%%%%%%%%%%%%%%%%%%%%%%%%%%%%%%%%%%%%%%%%%%%%%%%%%%%%%%%%%%%%%%

As mentioned in Sect. \ref{sec:Intr}, we selected three Terzan clusters for our study, namely Ter2, Ter4, and Ter5. These clusters exhibit a high probability of mutual interactions over their dynamical orbital histories. To reconstruct the evolutionary history of the mass and spatial parameters of the selected clusters at least approximately, we adopted the following initial models for each cluster individually. 

We assumed that the stellar mass distribution in our models spanned the range 0.08--100 M$_\odot$, following the Kroupa initial mass function \citep{Kroupa2001}. We also assumed for each cluster its own metallicity, which affects the stellar mass distribution and subsequent evolution. The models were initially brought into dynamical equilibrium assuming the King models \citep{King1966}, which are characterised by the half-mass radius r$_{\rm hm}$ and the central concentration parameter $W_0$. These three parameters (initial M, r$_{\rm hm}$, and $W_0$) were selected so that after 8 billion years of stellar and orbital evolution, the cluster system would have an M and r$_{\rm hm}$ comparable to observational data (see Sect. \ref{sec:mass-loss}).

Each particle in the simulation represented ten real stars of the same type, which allowed us to fit the cluster models more efficiently, particularly in terms of total mass and half-mass radius, to match their observed present-day values. This approximation in the particle number does not affect the system two-body relaxation time significantly. A detailed discussion of the effect of this reduction factor can be found in \cite{Taras2019} and in \cite{Ishchenko2024}. We refer to this setup as the {\tt common 3xGC run}, as all three Terzan clusters are modelled simultaneously. 

\begin{figure*}[tbh]
\centering
\includegraphics[width=0.99\linewidth]{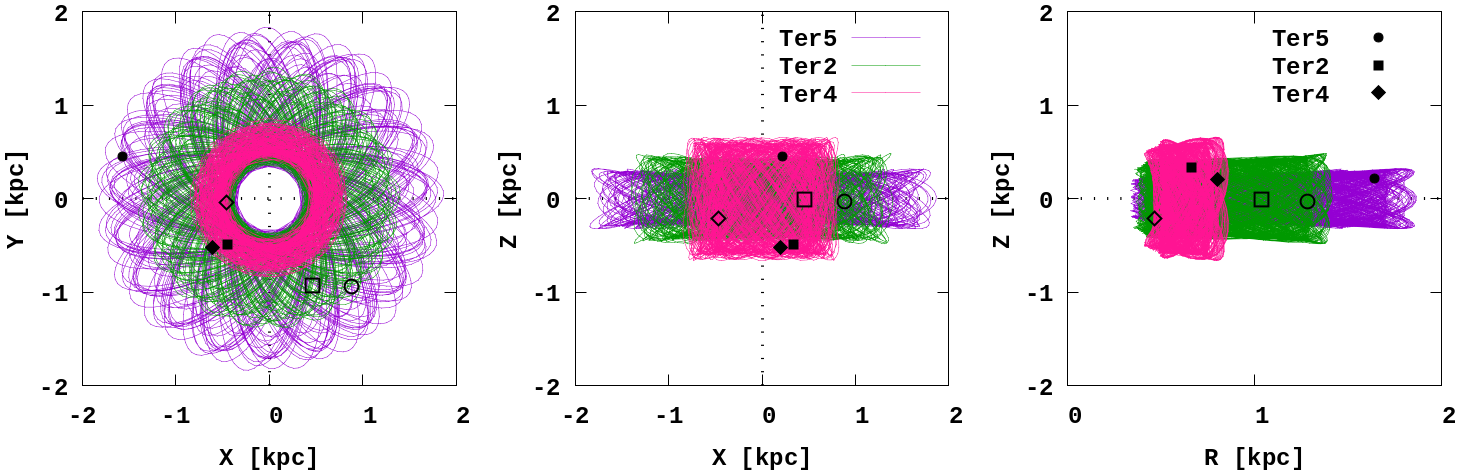}
\caption{Simultaneous orbital evolution of the Terzan clusters in the TVP {\tt 411321}. The orbital evolution is present in three planes, ($X-Y$), ($X-Z$), and ($R-Z$), where $R$ is the planar Galactocentric radius. The total integration time is an 8 Gyr look-back time. The black marks show the current positions, and empty marks show the positions at -8 Gyr.}
\label{fig:orb-3gc}
\end{figure*}

Table \ref{tab:init-param} summarises these physical parameters of the clusters at $T = -8$ Gyr. The initial positions and velocities for GC centres of mass used in the simulations correspond to their calculated stages 8 Gyr ago, obtained by backward integration of each cluster (treated as a point mass; these results were presented in detail in our \hyperlink{I23}{\color{blue}{Paper~I}}). The orbital evolution of the selected Terzan clusters in {\tt 411321} time-variable external potential is shown in Fig. \ref{fig:orb-3gc}. We also indicate the position for each GCs at -8 Gyr as empty marks in the plots. The effect of dynamical friction on the GC orbital evolution in the context of the point-mass approximation is discussed in App. \ref{app:orb-df}.  

%-------------------------------------------------------------------------%
\begin{table}[bpt]
\setlength{\tabcolsep}{4pt}
\centering
\caption{Initial physical characteristics at -8 Gyr for the Terzan clusters. %(252 rand seed).
}
\label{tab:init-param}
\begin{tabular}{lcccccc}
\hline
\hline 
GC &  $M$ & N & r$_{\rm hm}$ & $W_0$ & $Z_{\rm m}$ \\
 &  $10^{6}\rm\;M_{\odot}$ & & pc & & {$10^{-3}$} \\
\hline
\hline
Terzan 2  & 1.1 & 1 916 827 & 5.0 & 9.0 & 3.17 \\
Terzan 4  & 1.3 & 2 265 341 & 5.0 & 9.0 & 0.50 \\
Terzan 5  & 2.3 & 4 007 900 & 2.0 & 3.0 & 3.17 \\
\hline 
\end{tabular}
\vspace{6pt}
\end{table}
%-------------------------------------------------------------------------%

%-------------------------------------------------------------------------%
%-------------------------------------------------------------------------%

The numerical experiments were aimed to obtain well-fitted individual models for each GC by varying key parameters, including the total mass, half-mass radius, and King concentration parameter. The {\tt ref. runs}, in which full $N$-body simulations of each individual Terzan clusters were carried out separately, were conducted on the computer cluster of the Fesenkov Astrophysical Institute in Almaty, Kazakhstan and as on the LOTR cluster at the Main Astronomical Observatory of the National Academy of Sciences of Ukraine. Each run used a pair of NVIDIA RTX 4080 or RTX 4090 GPUs, achieving an average sustained performance of $\approx$21.2 Tflops per card. The {\tt common 3xGC run} was executed on the JUWELS Booster supercomputer at the Jülich Supercomputing Centre under the {\tt madnuc} project, using four nodes, each with a 4$\times$NVIDIA A100 GPU. The total modelling time took $\approx$5 months for the largest {\tt common 3xGC run}.

%%%%%%%%%%%%%%%%%%%%%%%%%%%%%%%%%%%%%%%%%%%%%%%%%%%%%%%%%%%%%%%%%%%%%
\section{Mass loss}\label{sec:mass-loss}
%%%%%%%%%%%%%%%%%%%%%%%%%%%%%%%%%%%%%%%%%%%%%%%%%%%%%%%%%%%%%%%%%%%

The long-term mass evolution of a GC is shaped by internal and external galactic dynamical processes. One significant source of cluster mass loss (especially in the first few hundred million years of evolution) is high-mass stellar evolution. As a result of the high-mass stars stellar evolution, we also observe the kick processes of black holes and neutron stars from the cluster. Clusters on eccentric orbits can experience stronger tidal forces, especially near their pericentre, which also enhances the stellar stripping and accelerates the subsequent mass loss \citep{Spitzer1987, Webb2014}. The evolving galactic potential can further amplify these effects by driving changes in orbital energy and angular momentum over time. The stripped stars often form extended tidal tails that trace the cluster orbit and even provide observable signatures of its dynamical history \citep{Habibi2013, Park2018, Arnold2025}. This is particularly relevant for central MW region clusters such as Ter5, Ter4, and Ter2, which follow compact eccentric orbits with pericentric distances of $\lesssim$0.5 kpc. Their proximity to the Galactic centre and repeated passages through these dense regions make them highly susceptible to tidal stripping, which offers valuable insight into mass loss and the formation of tidal debris.

\begin{figure*}[htb!]
\centering
\includegraphics[width=0.99\linewidth]{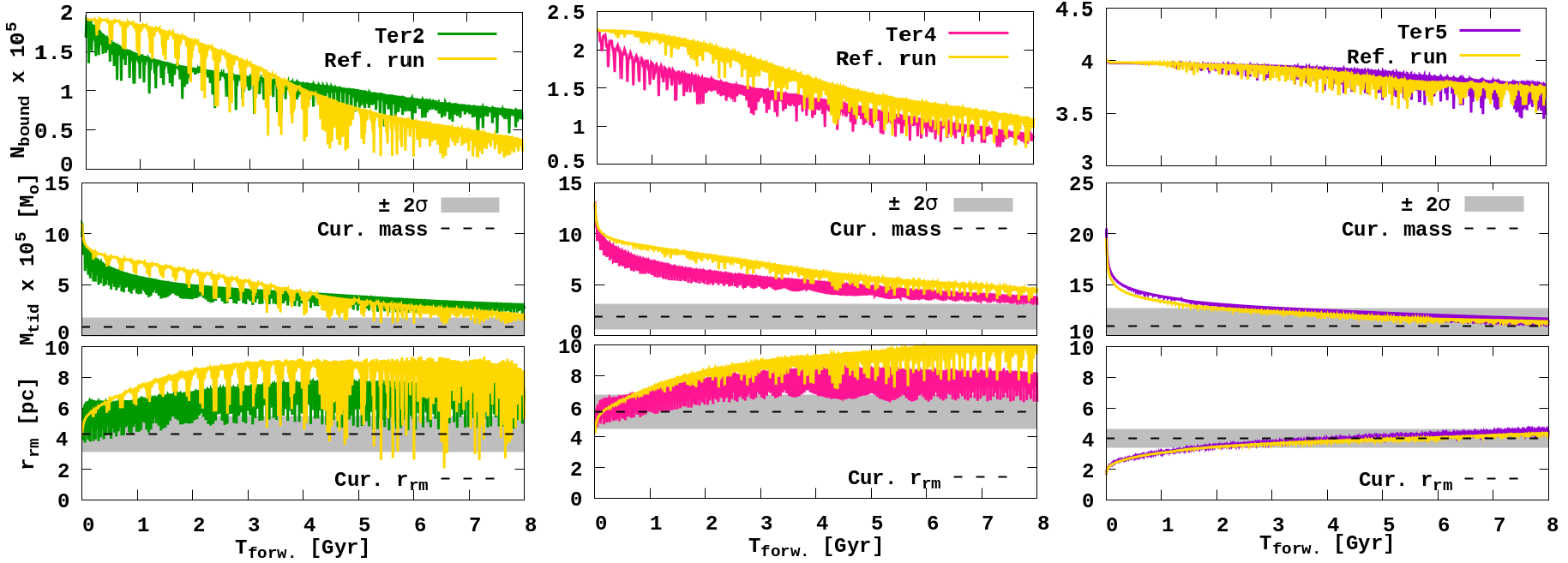}
\caption{Evolution of M$_{\rm tid}$ and r$_{\rm hm}$ for the Terzan clusters due to mass loss based on {\tt common 3xGC run} (coloured lines) and {\tt ref. run} (yellow lines). The dotted lines show the estimated current mass, $r_{\rm hm}$, and the grey zone shows the $\pm2\sigma$ observational uncertainty based on the catalogue by \cite{Baumgardt2021} for today.}
\label{fig:mtid-rhm}
\end{figure*}

In Fig.~\ref{fig:mtid-rhm} we present the evolution of the mass within a tidal radius and of the half-mass radii for Terzan clusters in more detail. The cluster tidal radius (also known as Jacobi or King radius) was calculated based on the numerical iteration of the $M_{\rm tid}$ and $r_{\rm tid}$ values according to the King equation,
\be
r_{\rm tid} = \left[ \frac{G \cdot M_{\rm tid}}{4\Omega^2-\kappa^2} \right]^{1/3}, 
\ee
where $G$, $M_{\rm tid}$, $\Omega$, and $\kappa$ are the gravitational constant, the tidal cluster mass, the circular and the epicyclic frequencies, respectively, of a near-circular orbit in the Galaxy at the GC current location (for more details, see \cite{King1962} and \cite{Ernst2011}). 

%-------------------------------------------------------------------------%
%-------------------------------------------------------------------------%

The grey bands in Fig. \ref{fig:mtid-rhm} represent the $\pm2\sigma$ uncertainties of the present-day tidal mass and half-mass radius estimates for the selected GCs, taken from the \cite{Baumgardt2021} catalogue. For Ter2 and Ter4, the model uncertainties for the final tidal mass are $\approx$29\%, whereas for Ter5, the uncertainty is only $\sim$7\%. This indicates that in our constructed initial models, the mass of Ter5 is determined most precisely. On the other hand, the masses of Ter2 and Ter4 are less well constrained.

The tidal stripping experienced by the GCs is also expected to lead to the formation of extended tidal tails. These tidal features are clearly present in our simulations. However, their detection in observations remains challenging, as their typical volume densities are very low ($\approx$0.01 M$_\odot$/pc$^3$), significantly below the average stellar mass density in the central Galaxy regions \citep{erik2009milky}. As a result, their identification likely requires full 6D phase-space information, which may become available with future data releases from the Gaia space mission.

The plots show that Ter5 lost about 45\% of its initial mass. For Ter2 and Ter4, the mass-loss evolution was quite similar, but by the present day, they already lost almost 75\% of their initial masses. To summarise, Ter5 loses no more than half of its initial mass, in contrast to the other two clusters, which undergo significantly stronger mass loss. Ter5 clearly exhibits the smallest fluctuations in its parameters throughout the evolution and provides the best match to its present-day properties. In contrast, the values of the parameters for Ter2 and Ter4 at the end of the simulation are systematically higher than the observed values and even lie outside their respective $\pm2\sigma$ uncertainty ranges. In addition, the initial models we assumed for the Terzan cluster family members are probably good enough for the purpose of studying their long-term mutual gravitational interactions. 

As initial conditions for these {\tt ref. runs} for each GC, we used exactly the same particle data as listed in Table \ref{tab:init-param}. Using these reference runs, we were able to examine the individual differences in the dynamical evolution of each Terzan GC. In Fig. \ref{fig:mtid-rhm} we present the dynamical mass-loss evolution for each cluster for the {\tt ref. runs} and the {\tt common 3xGC run}. The plots show that in general, the particle and mass loss, together with the half-mass radius evolution, are similar for all three systems, with some differences, especially for the lower-mass Ter2 and Ter4. In the case of the more massive Ter5, the {\tt common 3xGC run} shows practically the same mass-loss evolution as in the Ter5 {\tt ref. run}. 

To understand this mass-loss behaviour in different GCs on the cosmological timescale, we compared the mutual gravitational effect of the clusters in the {\tt common 3xGC run}. The most massive Ter5 clearly has the most significant gravitational perturbation acting on the other two Terzan systems. The effect of the low-mass Ter2 and Ter4 systems on Ter5 is significantly weaker. This effect is clear in the mass-loss and half-mass radius plots in Fig. \ref{fig:mtid-rhm}. This plot shows the combined effect of global gravitational perturbations, which act constantly, together with collisions (close passages) between individual Terzan systems. The effect of the global gravitational perturbation is visible almost immediately, after a few hundred million years from the beginning of the simulation, in the higher mass-loss rate compared with the corresponding {\tt ref. runs}. By the end of the simulation, the mass loss for the individual {\tt ref. runs} and for the {\tt common 3xGC} model becomes very similar. This effect is naturally stronger for Ter2 and Ter4.

We refer to App. \ref{app:rem--den-distr} for details about the internal dynamic distribution of massive star remnants for Terzan clusters after an evolution of 8 Gyr. We present the analysis as from {\tt common 3xGC run} and as {\tt ref. run} in this appendix.

%%%%%%%%%%%%%%%%%%%%%%%%%%%%%%%%%%%%%%%%%%%%%%%%%%%%%%%%%%%%%%%%%%%%%
\section{Terzan 2, 4, and 5 collision tracking on a cosmological timescale}\label{sec:gc-timescale}
%%%%%%%%%%%%%%%%%%%%%%%%%%%%%%%%%%%%%%%%%%%%%%%%%%%%%%%%%%%%%%%%%%%%

Throughout the simulation, we monitored the possible close approaches between GCs at fixed distances shorter than $dR=100$ pc. In Table \ref{tab:stat-coll-pair} we list the GC pairs involved in these events, along with the time of the encounter. We also present the minimum separation and the relative velocity between their centres of mass together with the corresponding values of their tidal radii at that moment.

%-------------------------------------------------------------------------%
\begin{table}[tbp]
\centering
\caption{Physical parameters for the collisional pairs in the {\tt common 3xGC run}.}
\label{tab:stat-coll-pair}
\begin{tabular}{ccccccc}
\hline
\hline 
& GC pair & T$_{\rm coll}$ & dR$_{\rm pair}$ & dV$_{\rm pair}$ & r$_{\rm tid}$, GC$_1$ & r$_{\rm tid}$, GC$_2$ \\
 & & Gyr & pc & km/s & pc & pc \\
%(1) & (2) & (3) & (4) & (5) & (6) & (7) & (8) & (9) & (10) & (11) & (12)  \\
\hline
\hline
 & Ter2--Ter4 & 0.22 & 72 & 283 & 17 & 21 \\
 & Ter2--Ter4 & 0.85 & 72 & 289 & 17 & 22 \\
 & Ter2--Ter4 & 1.83 & 60 & 194 &  21 & 33 \\
{\tt A} & {\bf Ter2--Ter4} & 1.89 & 42 & 233 & 18 & 29 \\
& Ter2--Ter4 & 2.04 & 60 & 317 & 12 & 18 \\
& {\bf Ter2--Ter4} & 2.19 & 10 & 320 & 12 & 17 \\
 & {\bf Ter2--Ter4} & 2.42 & 18 & 213 & 14 & 19 \\
 & Ter2--Ter4 & 3.24 & 70 & 198 & 19 & 31 \\
 & Ter2--Ter4 & 4.09 & 60 & 276 & 15 & 19 \\
 & Ter2--Ter4 & 4.32 & 89 & 226 & 18 & 23 \\
 & Ter2--Ter4 & 4.53 & 89 & 214 & 18 & 32 \\
 & Ter2--Ter4 & 6.65 & 45 & 205 & 14 & 20 \\
 & Ter2--Ter4 & 6.90 & 77 & 234 & 17 & 27 \\
 & \textbf{Ter2--Ter4} & 7.57 & 25 & 206 & 20 & 27 \\
\hline
 & Ter2--Ter5 & 4.66 & 54 & 321 & 19 & 24 \\
 & Ter2--Ter5 & 6.00 & 51 & 318 & 21 & 26 \\
 & Ter2--Ter5 & 7.31 & 68 & 187 & 22 & 30 \\
\hline
 & Ter4--Ter5 & 0.56 & 54 & 241 & 18 & 17 \\
 & Ter4--Ter5 & 1.19 & 48 & 187 & 17 & 18 \\
 & {\bf Ter4--Ter5} & 2.12 & 30 & 199 & 20 & 19 \\
 & Ter4--Ter5 & 3.36 & 74 & 155 & 18 & 18 \\
 & Ter4--Ter5 & 3.81 & 62 & 175 & 20 & 22 \\
 & Ter4--Ter5 & 5.82 & 83 & 151 & 16 & 18 \\
 & {\bf Ter4--Ter5} & 7.24 & 22 & 202 & 24 & 24 \\
\hline 
\end{tabular}
\tablefoot{In column T$_{\rm coll}$, 0.00 means -8 Gyr ago, \textbf{forward} run.}
\end{table}
%-------------------------------------------------------------------------%

The most significant of the recorded encounters are four close collisions between Ter2 and Ter4 and two collisions between Ter4 and Ter5. In these cases, the sum of the tidal cluster radii exceeds their minimum separation. A particularly notable event is the encounter between Ter2 and Ter4, during which the clusters approached to within 10 pc, while their tidal radii were 12 and 17 pc, respectively. In this case, however, the relative fly-by velocity between clusters is also one of the highest for this pair ($\sim$320 km/s). As expected, Ter4 remains the most dynamically active cluster in terms of close encounters and interactions with the other two systems.

\begin{figure}[tb!]
\centering
\includegraphics[width=0.95\linewidth]{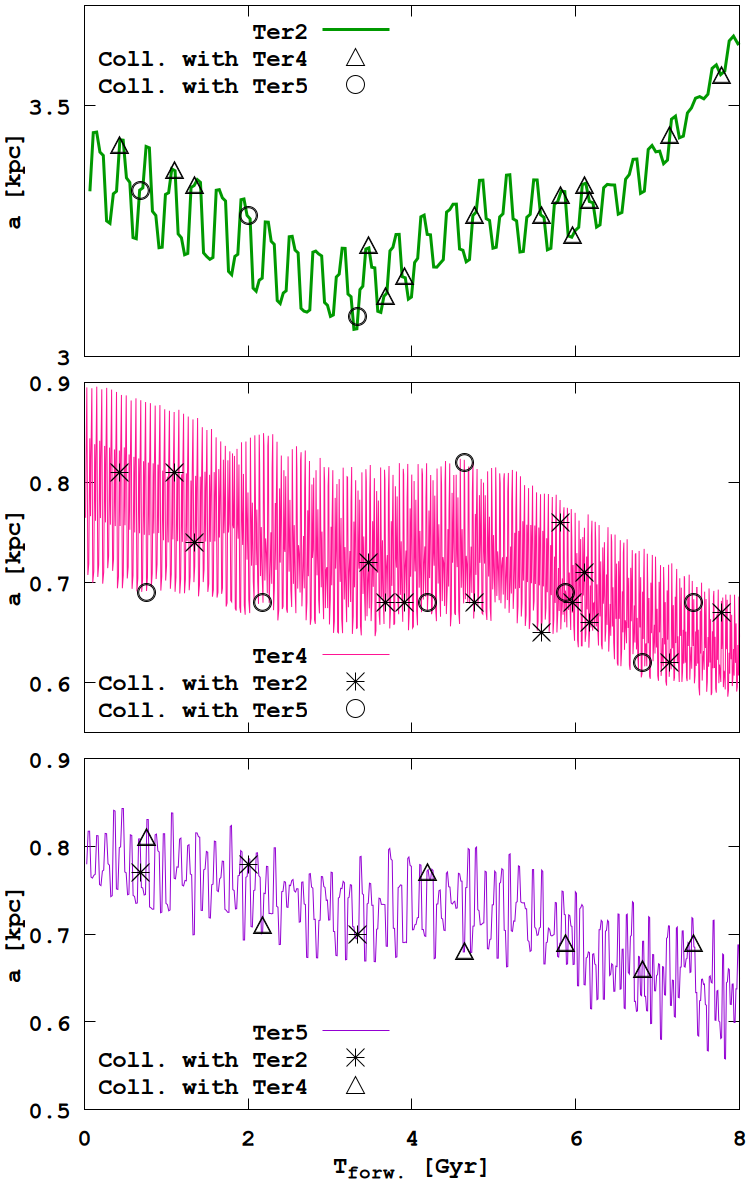}
\caption{Evolution of the GC semi-major axes during the simulation with marked collision events in the {\tt common 3xGC run}.}
\label{fig:a-ters}
\end{figure}

In Fig. \ref{fig:a-ters} we present the evolution of the semi-major axis, $a$, of each cluster over a simulation of 8 Gyr and mark the collision events (according to Table \ref{tab:stat-coll-pair}). The encounters are distributed relatively uniformly in time and in semi-major axis, indicating no clear dependence of the close interaction probability on orbital phases.

\begin{figure*}[tb!]
\centering
\includegraphics[width=0.45\linewidth]{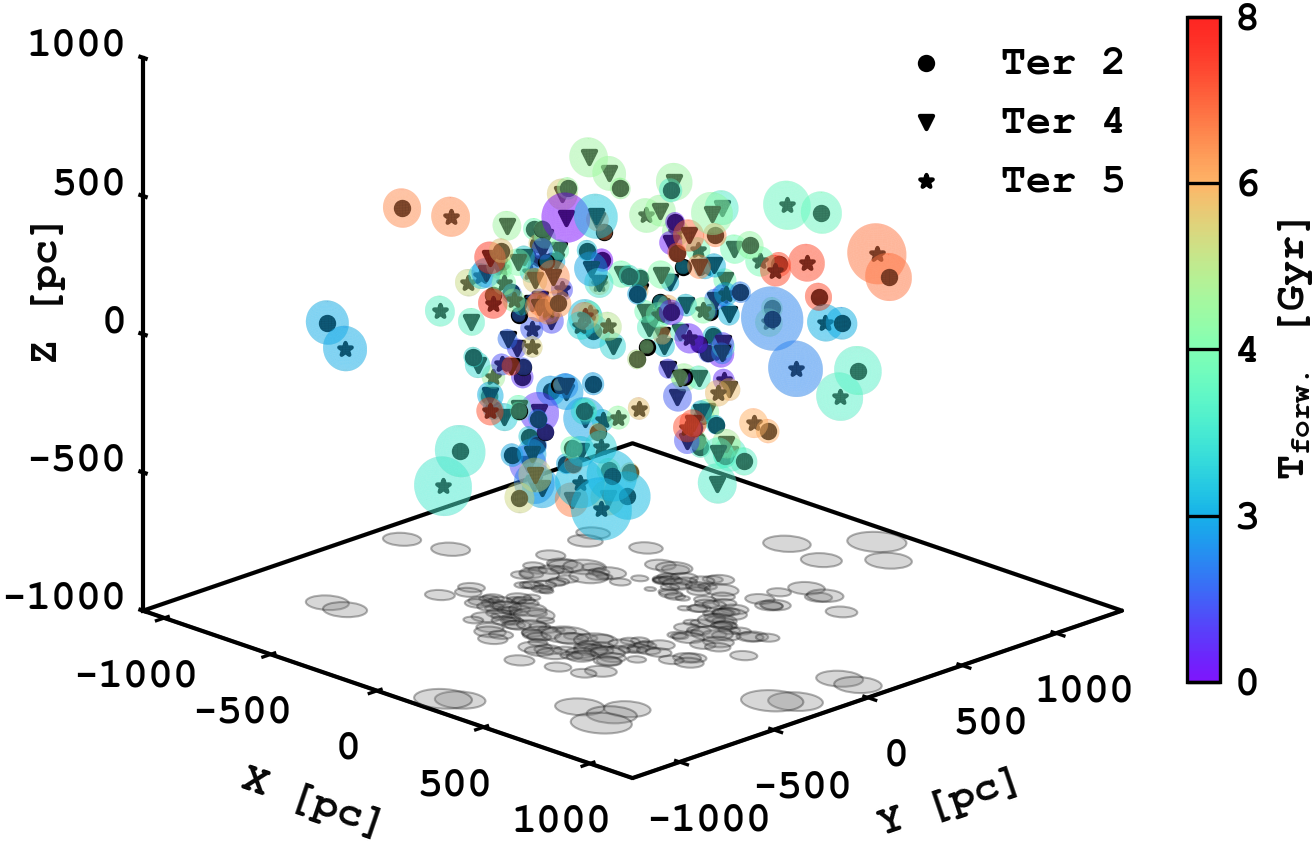}
\includegraphics[width=0.42\linewidth]{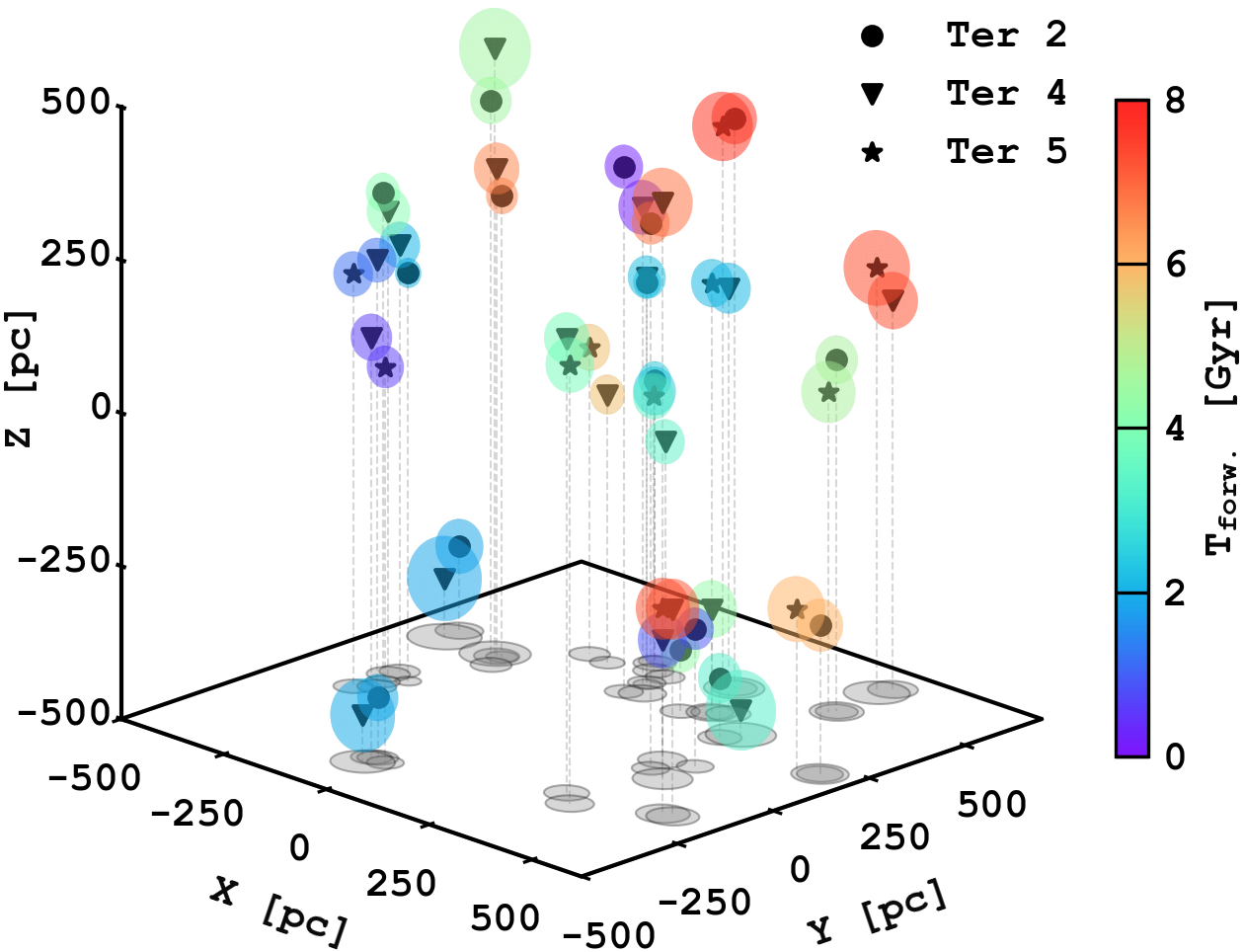}
\caption{Collisions of the Terzan clusters in 3D projections for different $dR$ conditions. Left: $dR=200$ pc. Right: $dR=100$ pc.}
\label{fig:3d-coll}
\end{figure*}

The right panel of Fig. \ref{fig:3d-coll} shows the collision pairs from Table \ref{tab:stat-coll-pair} in a 3D space projection, with a colour-coding by time. The coloured spheres represent the doubled tidal radii of the two clusters at the moment of the encounter. The left panel displays the position of cluster pairs that approached each other within 200 pc. A sharp increase in the number of such pairs is evident for $dR=200$ pc. In the \textit{X-Y} projection, they form a ring-like structure around the Galactic centre. This effect was already  observed in  \hyperlink{I232}{\color{blue}{Paper~III}} (see Fig. 8).

%-------------------------------------------------------------------------%
%-------------------------------------------------------------------------%

%-------------------------------------------------------------------------%
%-------------------------------------------------------------------------%

%%%%%%%%%%%%%%%%%%%%%%%%%%%%%%%%%%%%%%%%%%%%%%%%%%%%%%%%%%%%%%%%%%%%%
\section{Gravitational potential perturbations caused by the mutual global effect of Terzan 2, 4, and 5}\label{sec:tail}
%%%%%%%%%%%%%%%%%%%%%%%%%%%%%%%%%%%%%%%%%%%%%%%%%%%%%%%%%%%%%%%%%%%%

Based on our previous analysis of the {\tt common 3xGC run} (three GCs modelled together), we already detected some common effects of three separate GCs in the time-evolved external gravitational field of the Galaxy. For a detailed analysis of this common gravitational effect of the GCs on each other, we decided to carry out a new set of separate $N$-body runs in which we modelled the dynamical evolution of each Terzan GC individually in the same external MW-like potential.  

\begin{figure*}[tb!]
\centering
\includegraphics[width=0.46\linewidth]{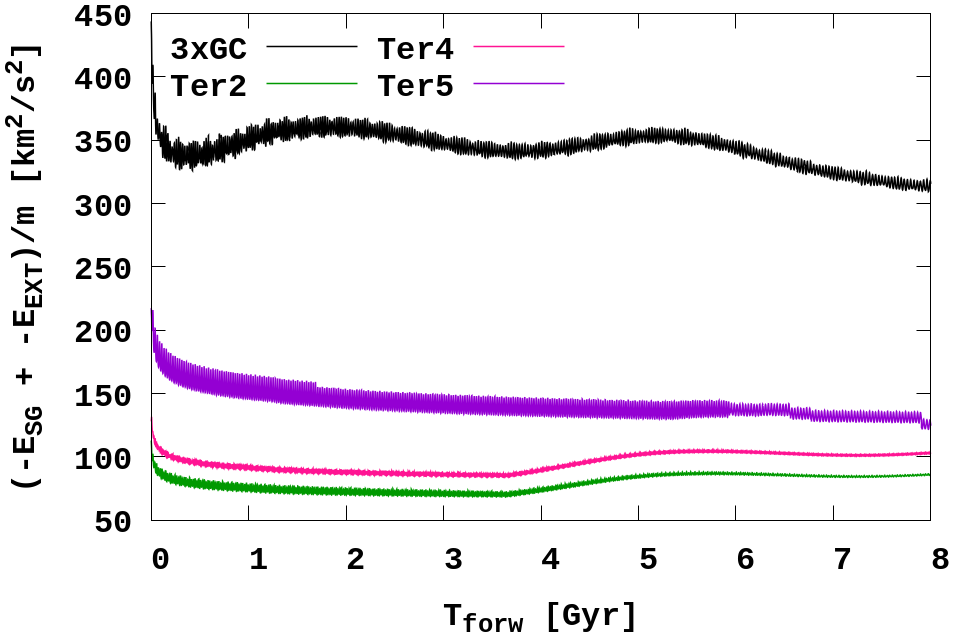}
\includegraphics[width=0.48\linewidth]{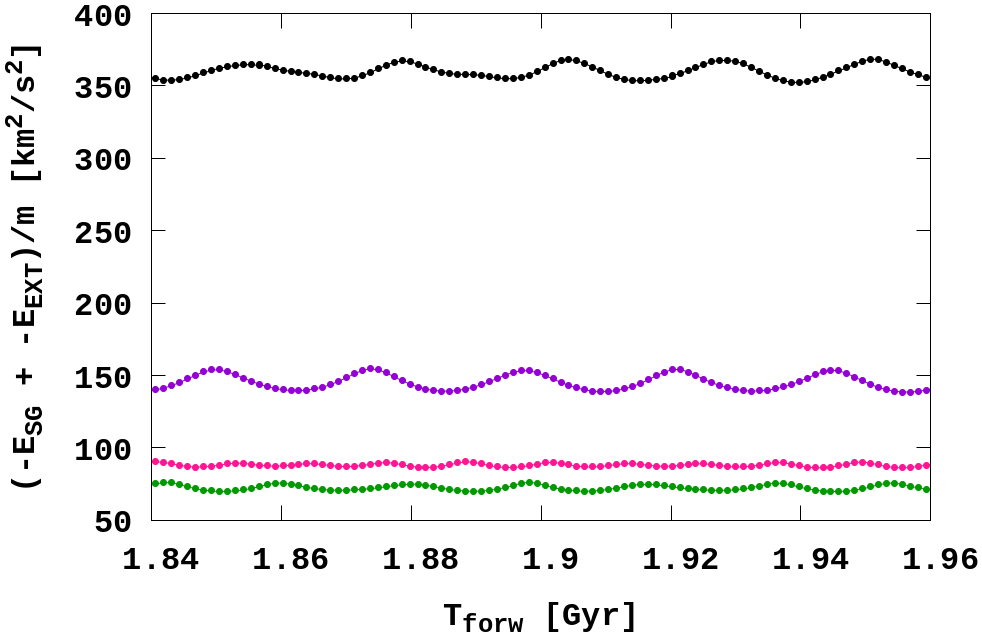}
\caption{Global potential energy evolution of the cluster systems for the {\tt common 3xGC run} and the individual {\tt ref. runs}. Left: Full-time evolution of the models up to 8 Gyr. Right: Same energy evolution with a very detailed output near the close passage marked {\tt A} in Table \ref{tab:stat-coll-pair}.}
\label{fig:energy-all}
\end{figure*}

To quantitatively check this effect, we analysed the behaviour of the energy outputs for the {\tt ref. runs} and the {\tt common 3xGC run}in high detail. We present these results in the left panel of Fig. \ref{fig:energy-all}. This plot shows the total potential energy of the systems, namely the sum of particle self-gravity and their energy in the external gravitational field. In the {\tt common 3xGC run} results (black line), the energy perturbation level is significant because the three GCs collectively affect each other gravitationally. This perturbation is present throughout the entire simulation time. To obtain a more detailed view of this cyclic perturbation, we present the very detailed time output of these energies during one of the Ter2 -- Ter4 close-passage events (marked {\tt A} in Table \ref{tab:stat-coll-pair}) in the right panel of Fig. \ref{fig:energy-all}. In addition to the higher potential energy level in the case of the {\tt common 3xGC run}, we also detect low-level short-timescale periodic perturbations of $\sim$200 Myr that reflect the mutual interactions between the three GCs.             

%-------------------------------------------------------------------------%
%-------------------------------------------------------------------------%

To demonstrate the collective behaviour of the GC system in the case of the {\tt common 3xGC run,} we also analysed the ellipsoidal shape of the GC main stellar body in detail. To determine the semi-major axis, we used the well-tested algorithm {\tt triax}. This method is described in detail in Appendix C in \cite{Regaly2023}. The {\tt triax} method is based on the well-known Jacobi eigenvalue matrix algorithm\footnote{Jacobi eigenvalue matrix algorithm: \\~\url{ https://en.wikipedia.org/wiki/Jacobi_eigenvalue_algorithm}}. As a result of the method, we obtained the ratios of the semi-major axis $b/a$ and $c/a$ and as a final parameter: the evolution of the triaxiality, defined as T = (a$^2$ -- b$^2$)/(a$^2$ -- c$^2$). 

\begin{figure}[tb!]
\centering
\includegraphics[width=0.99\linewidth]{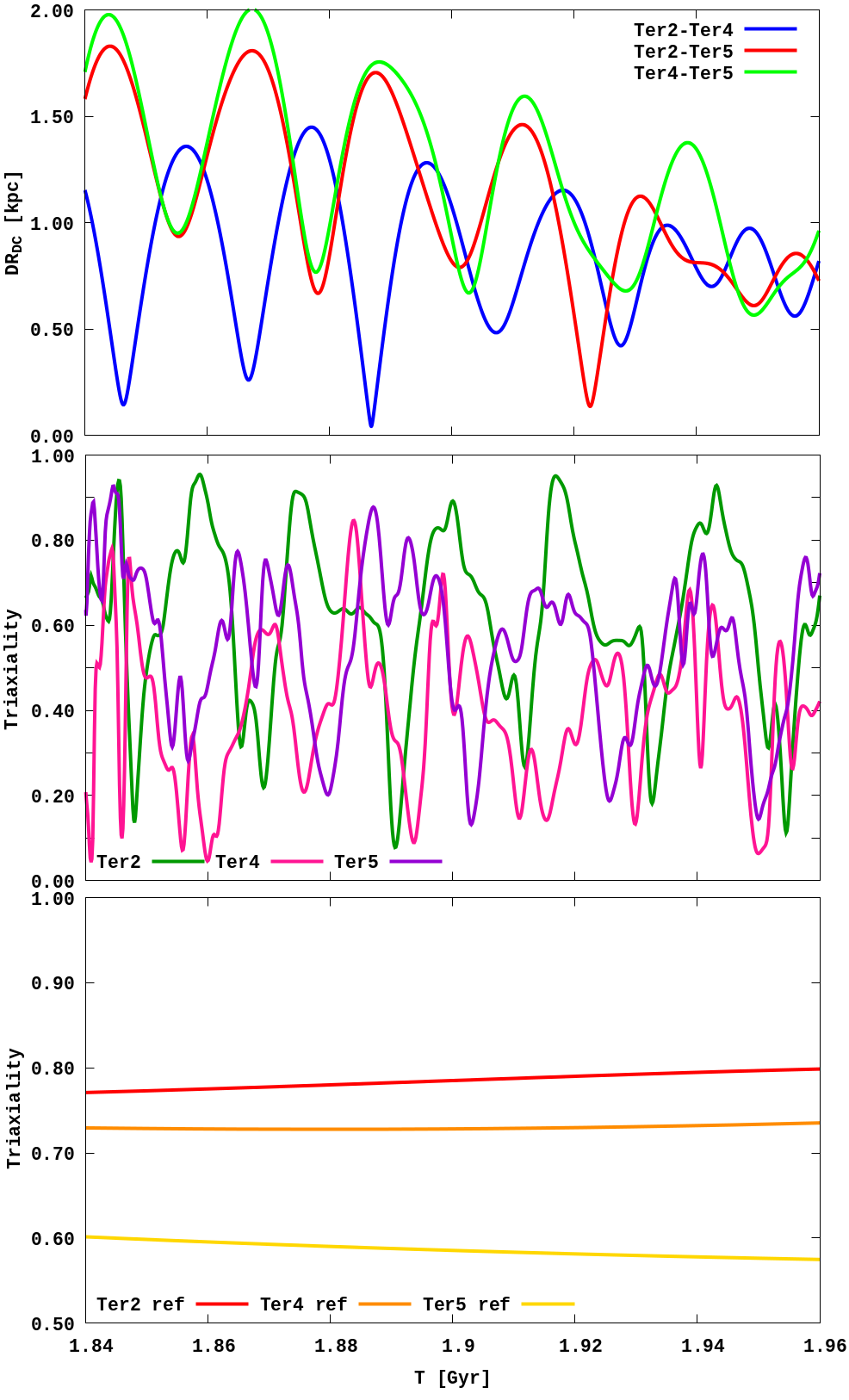}
\caption{Time evolution of the mutual distance between GC pairs and the axial ratios for the {\tt common 3xGC run} and {\tt ref. runs}. The middle and bottom panels show the triaxiality. The top and middle panels correspond to the {\tt common 3xGC run}, and the bottom panel shows the {\tt ref. runs} for the Terzan clusters.}
\label{fig:triax}
\end{figure}

Figure~\ref{fig:triax} shows the time evolution of the cluster {\tt triax as the T} parameter for {\tt common 3xGC run} and {\tt ref. runs} for highly detailed outputs for the {\tt A} case of closed passage. In the top panel, we present the time evolution of the mutual distances between the pairs of clusters, which is calculated as 
$$
{\rm DR}_{\rm DC} = \sqrt{(X_{\rm GC,i} - X_{\rm GC,j})^{2} + (Y_{\rm GC,i} - Y_{\rm GC,j})^{2} + (Z_{\rm GC,i} - Z_{\rm GC,j})^{2}}.
$$ 

In the middle panel of Fig. \ref{fig:triax} we present the triaxial parameter definition for the {\tt common 3xGC run}. In the bottom plot, we present the same {\tt triax} evolution for {\tt ref. runs}. The figures clearly show the differences between the two types of runs. The {\tt ref. runs} evolve smoothly for the selected time interval, without significant changes in the {\tt triax} parameter at one level for Ter2 in 0.78, for Ter4 in 0.74, and for Ter5 in 0.60. For the {\tt common 3xGC run}, the behaviour is completely different: there are strong and short fluctuations of {\tt T} caused by the changing mutual separation between the clusters. The overall variability of {\tt T} is very strongly correlated with the DR$_{\rm DC}$ in the top panel. In \ref{fig:triax-a-b} we also provide the evolution $b/a$ and $c/a$ axes separately for the {\tt common 3xGC run} and {\tt ref. runs}.

%-------------------------------------------------------------------------%
%-------------------------------------------------------------------------%

%%%%%%%%%%%%%%%%%%%%%%%%%%%%%%%%%%%%%%%%%%%%%%%%%%%%%%%%%%%%%%%%%%%%%%%%%%%%%%%
\section{Conclusions}\label{sec:con}
%%%%%%%%%%%%%%%%%%%%%%%%%%%%%%%%%%%%%%%%%%%%%%%%%%%%%%%%%%%%%%%%%%%%%%%%%%%%%%%

The evolution of our GC subsystem remains one of the most actively discussed topics in contemporary Galactic astrophysics. GCs provide crucial insights into the early stages of Galaxy formation, including past accretion events and the subsequent dynamical evolution of the system.

After performing a set of high-resolution dynamical $N$-body simulations of the three Galactic centre GCs Ter2, Ter4, and Ter5, we analysed the individual and mutual evolution of these systems in detail and identified a list of pairwise collisions (close encounters; see Table \ref{tab:stat-coll-pair}). These collisions persisted during the whole modelling time of 8 Gyr. In a few cases, the separation between the cluster pairs became smaller than the sum of their r$_{\rm tid}$. During these events, the relative velocities of the clusters were quite high, $\sim$200--300 km/s. These high velocities prevent the possible formation of gravitationally bound Terzan cluster pairs. 

An interesting finding of our simulations is the slightly different mass-loss behaviour. The individual single Terzan full $N$-body simulations ({\tt ref. runs}) show a significantly lower initial mass loss (particle loss) during the first few hundred million years compared to the {\tt common 3xGC} model, in which all three clusters were simulated simultaneously (see Fig. \ref{fig:mtid-rhm}). At the same time, the potential energy differences between the {\tt common 3xGC} and {\tt ref. runs} were significant (see Fig. \ref{fig:energy-all}). 

The collective behaviour of the GC system in the {\tt common 3xGC run} was analysed, revealing a clear effect of mutual interactions on the evolution of the cluster shapes. The application of the {\tt triax} algorithm demonstrated strong correlations between the axis ratios and the mutual distances of the cluster pairs, especially for Ter2, which evolved from nearly spherical to distinctly prolate ($b/a$ and $c/a$ $\sim$0.75--0.80), and for Ter5, which showed similar but less pronounced changes (see Fig. \ref{fig:triax}). In contrast, Ter4 remained relatively stable. In the {\tt ref. run}, however, the clusters preserved almost spherical shapes (with axis ratios $\gtrsim$ 0.95), highlighting that the triaxial deformations observed in the {\tt common 3xGC} case are driven by mutual gravitational interactions among the modelled Terzan clusters.

The high-mass stellar remnant normalised cumulative distributions for these two groups of runs also showed a more compact number distribution result for the {\tt common 3xGC run} today. This effect is especially clearly manifested for the low-mass systems Ter2 and Ter4 (see App.~\ref{app:rem--den-distr}). 

We conclude based on these findings that the full group of GCs within this zone of a few kiloparsec should preferably be included in the modelling during the dynamical evolution of Galactic central-region clusters. This complex modelling is especially important in the case of low-mass systems because their high-mass neighbours, for example, Ter5 or Liller~1, which have current masses of more than 10$^6$ M$_\odot$, can affect their internal dynamical and shape behaviour significantly (see Fig. \ref{fig:mtid-rhm} and Fig. \ref{fig:triax}).  

Deviations in the GC shapes from spherical symmetry are commonly attributed either to internal mechanisms, such as rotation, or to external effects, primarily tidal forces exerted by the surrounding environment. In support of the latter, previous studies have reported systematic elongation of clusters located near the Galactic bulge, with their major axes preferentially aligned towards the Galactic centre \citep{Chen2010}. This suggests that the external tidal field can play a significant role in shaping the GC morphology in dense environments. However, observational constraints on the shapes of our chosen GCs remain absent in the literature. Therefore, while the effect of neighbouring massive clusters and the bulge potential is physically plausible, its quantitative effect on the morphology of individual systems is still not well constrained.

In our current modelling, we restricted the simulated clusters to the list of systems that survived the early turbulent dynamical history of the Galactic centre formation and evolution. These surviving systems are clearly less affected by global common gravitational interactions than the GC systems that were already dissolved during the early Galaxy evolution on cosmological timescales. The first attempt at systematic direct $N$-body modelling of such already dissolved GC systems in a realistic time-evolving MW potential was presented in our previous paper, \cite{Dana2026}. In the follow-up studies, we plan to extend this combined evolutionary approach to the global system of dissolved and surviving Galactic centre GCs. The question of individual GC survivability is a complex multi-parameter problem. According to our current understanding, it requires not only a description of the internal dynamics of the GC, but also a correct account of interactions with other nearby systems.    

%%%%%%%%%%%%%%%%%%%%%%%%%%%%%%%%%%%%%%%%%%%%%%%%%%%%%%%%%%%%%%%%%%%%%
\begin{acknowledgements}
%%%%%%%%%%%%%%%%%%%%%%%%%%%%%%%%%%%%%%%%%%%%%%%%%%%%%%%%%%%%%%%%%%%%%
The authors thank the anonymous referee for a very constructive report and suggestions that helped significantly improve the quality of the manuscript.

The work was carried out with the support of the Ministry of Science and Higher Education of the Republic of Kazakhstan within the framework of Project No AP26100669, ``Galactic Archaeology of Gaia-Enceladus Globular Clusters as Indicators of the Formation History of the Milky Way Disk''.

MI and PB thank the support from the special program of the Polish Academy of Sciences and the U.S. National Academy of Sciences under the Long-term program to support Ukrainian research teams, grant No.~PAN.BFB.S.BWZ.329.022.2023.

We also gratefully acknowledge the Polish high-performance computing infrastructure PLGrid (HPC Center: ACK Cyfronet AGH) for providing computer facilities and support within computational grant No.~PLG/2026/019243.

\end{acknowledgements}

%%%%%%%%%%%%%%%%%%%%%%%%%%%%%%%%%%%%%%%%%%%%%%%%%%%%%%%%%%%%%%%%%%%%%
\bibliographystyle{bibtex/aa}  % style aa.bst
\bibliography{bibtex/sources.bib}   % your references Yourfile.bib
%%%%%%%%%%%%%%%%%%%%%%%%%%%%%%%%%%%%%%%%%%%%%%%%%%%%%%%%%%%%%%%%%%%%%

%%%%%%%%%%%%%%%%%%%%%%%%%%%%%%%%%%%%%%%%%%%%%%%%%%%%%%%%%%%%%%%%%%%%%
\begin{appendix}
\onecolumn
%%%%%%%%%%%%%%%%%%%%%%%%%%%%%%%%%%%%%%%%%%%%%%%%%%%%%%%%%%%%%%%%%%%%%%%%%%%%%%%
\section{Orbital evolution of Terzan 2, 4, and 5 in the TVP with dynamical friction}\label{app:orb-df}
%%%%%%%%%%%%%%%%%%%%%%%%%%%%%%%%%%%%%%%%%%%%%%%%%%%%%%%%%%%%%%%%%%%%%%%%%%%%%%%

Here, we include additional computations aimed at quantifying the impact of dynamical friction (DF) on the determination of GCs' orbital parameters. In an astrophysical framework, DF describes the gradual deceleration experienced by a massive object moving through a background of field stars, caused by the cumulative effect of gravitational interactions with surrounding particles. This phenomenon was first identified by Chandrasekhar and von Neumann in their seminal papers \citep{CN1942, CN1943}. Subsequently, Chandrasekhar provided a more rigorous and quantitative formulation of the theory in later works \citep{C1943a, C1943b}. The resulting acceleration due to DF acting on a GC can be expressed as \citep{Binney2008}:
\begin{equation}
\frac{{\rm d} {\bf V}_{\rm GC} }{\rm dt} = -\frac{4\pi G^2\rho M_{\rm GC}}{V_{\rm GC}^3} \; \chi \cdot 
        \ln\Lambda \cdot {\bf V}_{\rm GC} 
    \quad \mathrm{with}\quad 
    \chi=\frac{\rho(<V_{\rm GC})}{\rho}.
    \label{dynfric}
\end{equation}

In general, the function $\chi$ and the Coulomb logarithm $\Lambda$ are functions of the velocity of the massive body, as well as of the physical characteristics of the surrounding medium. A detailed discussion of the adopted dynamical friction formalism is provided in \cite{Just2011}. Motivated by the results presented in that study, we assume constant values of $\ln\Lambda$ = 5 and a $\chi$ = 0.5 throughout our calculations.

\begin{figure*}[tbh]
\centering
\includegraphics[width=0.99\linewidth]{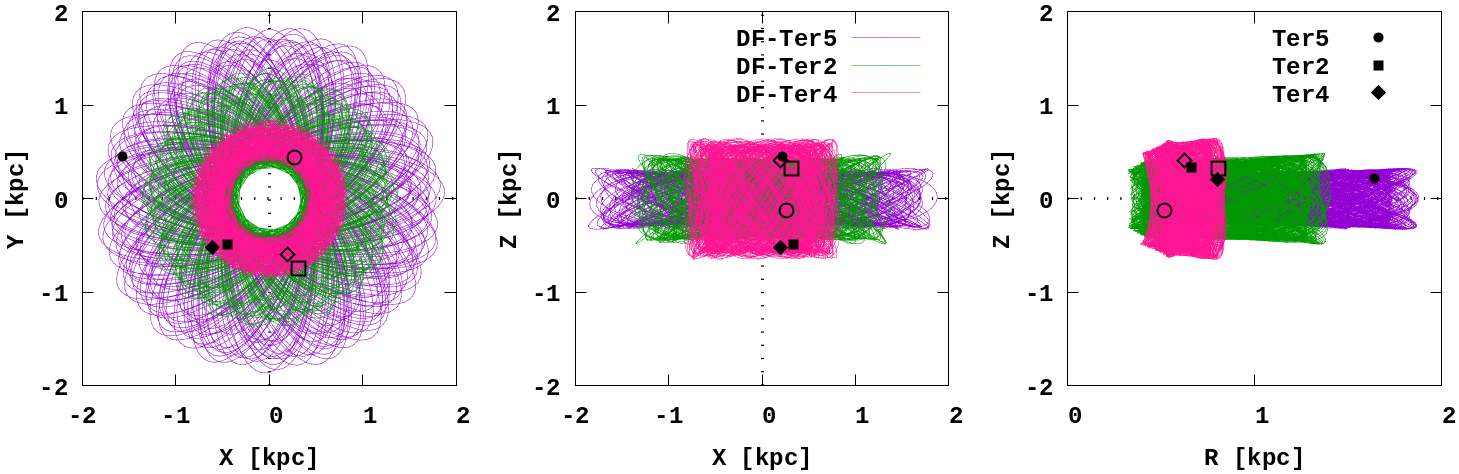}
\caption{Orbital evolution of the Terzan clusters integrated simultaneously in the TVP external potential {\tt 411321}, including DF. The trajectories are shown in the ($X-Y$), ($X-Z$), and ($R-Z$) planes, where $R$ is the planar Galactocentric radius. The total integration time corresponds to a look-back time of 8 Gyr. Black markers indicate the current positions, while non-filled markers indicate the positions at -8 Gyr.}
\label{fig:orb-3gc-df}
\end{figure*}

\begin{figure}[tbh]
\centering
\includegraphics[width=0.44\linewidth]{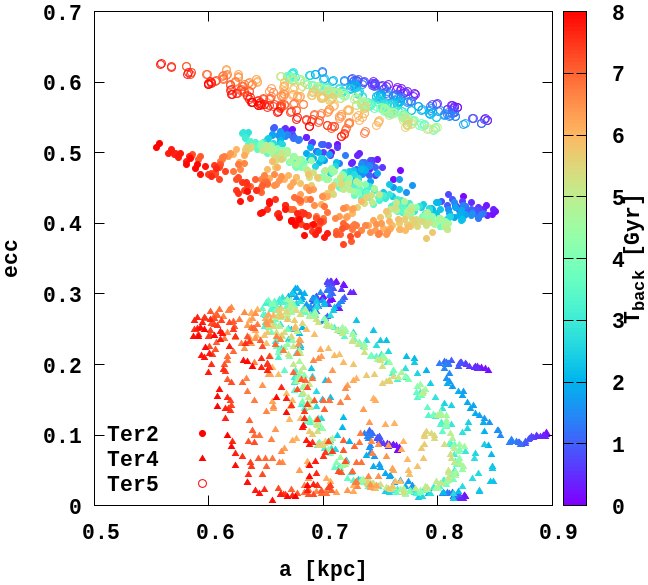}
\includegraphics[width=0.44\linewidth]{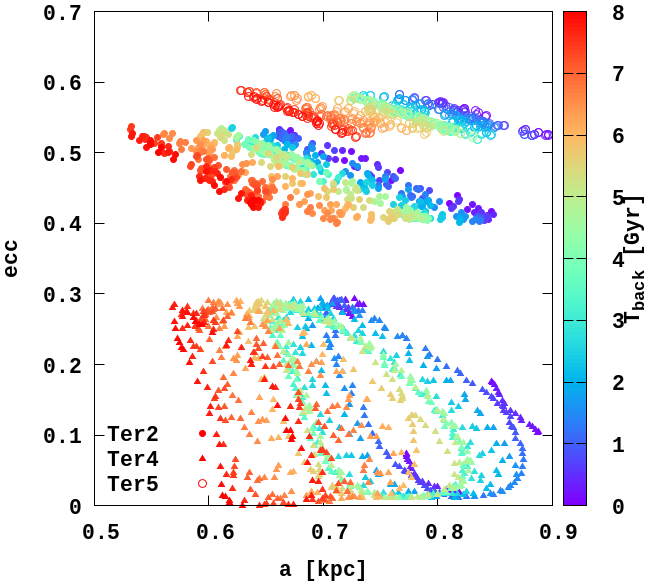}
\caption{Evolution of \textit{a} and {\tt ecc} for the Terzan clusters integrated simultaneously in the TVP external potential {\tt 411321}. Left: without DF. Right: with DF.}
\label{fig:a-ecc-df}
\end{figure}

In Fig. \ref{fig:orb-3gc-df}, we present the evolution of the orbits of our three GCs over 8 Gyr of integration time for models calculated with a dynamical friction algorithm. In the body of the main paper, this figure can be compared with Fig. \ref{fig:orb-3gc}. In Fig. \ref{fig:a-ecc-df}, we also show the evolution of the Terzan clusters' orbital parameters \textbf{a} and {\tt ecc} with and without DF. In our opinion, applying dynamical friction in its present form does not significantly change the Terzan cluster orbits on the global phase-space diagrams.

%-------------------------------------------------------------------------%
%-------------------------------------------------------------------------%

%-------------------------------------------------------------------------%
%-------------------------------------------------------------------------%

%%%%%%%%%%%%%%%%%%%%%%%%%%%%%%%%%%%%%%%%%%%%%%%%%%%%%%%%%%%%%%%%%%%%%%%%%%%%%%%
\section{Evolution of $b/a$ and $c/a$ axial ratios}\label{app:triax}
%%%%%%%%%%%%%%%%%%%%%%%%%%%%%%%%%%%%%%%%%%%%%%%%%%%%%%%%%%%%%%%%%%%%%%%%%%%%%%%

\begin{figure*}[tbh]
\centering
\includegraphics[width=0.95\linewidth]{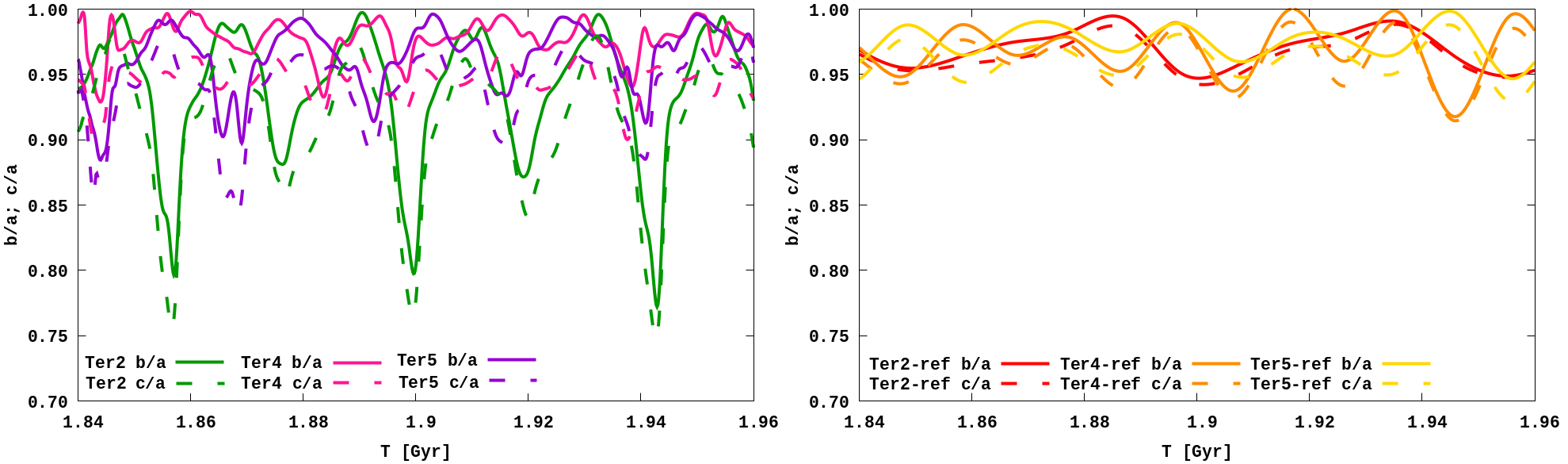}
\caption{Time evolution of the axial ratios, $b/a$ and $c/a$, for the {\tt common 3xGC run} and the {\tt ref. runs}.}
\label{fig:triax-a-b}
\end{figure*}

In Fig. \ref{fig:triax-a-b}, we show the evolution of the axial ratios, demonstrating strong correlations between the $b/a$ and $c/a$ ratios for the {\tt common 3xGC run} and the {\tt ref. runs}. The systems evolve from nearly spherical to distinctly prolate configurations ($b/a$ and $c/a$ $\sim$0.75--0.80), especially in the case of Ter2. Ter5 exhibits similar but less pronounced changes. In contrast, Ter4 remains relatively stable. In the {\tt ref. run}, however, the clusters retain nearly spherical shapes (axis ratios $\gtrsim$ 0.95), highlighting that the triaxial deformations observed in the {\tt common 3xGC} case are driven by mutual gravitational interactions among the Terzan clusters. It should be noted that the evolution of the $b/a$ and $c/a$ axes for Ter4 requires further modelling and research.

%-------------------------------------------------------------------------%
%-------------------------------------------------------------------------%

%%%%%%%%%%%%%%%%%%%%%%%%%%%%%%%%%%%%%%%%%%%%%%%%%%%%%%%%%%%%%%%%%%%%%%%%%%%%%%%
\section{High-mass stellar remnant distribution for the inner Terzan cluster regions}\label{app:rem--den-distr}
%%%%%%%%%%%%%%%%%%%%%%%%%%%%%%%%%%%%%%%%%%%%%%%%%%%%%%%%%%%%%%%%%%%%%%%%%%%%%%%

We focused in particular on compact stellar remnants located in the central region (inner 30 pc). By employing detailed stellar evolution modules within our $N$-body framework, we traced the internal kinematics and long-term orbital behaviour of remnants, primarily white dwarfs (WDs), neutron stars (NSs), and black holes (BHs; \citealt{Kamlah2022a}). The most dynamically impactful stage in the lives of massive stars occurs during their terminal supernova explosions. Such energetic events generally impart fallback-scaled natal kicks with random orientations to the newly formed remnants \citep{Banerjee2020}. 

\begin{figure*}[tbh]
\centering
\includegraphics[width=0.99\linewidth]{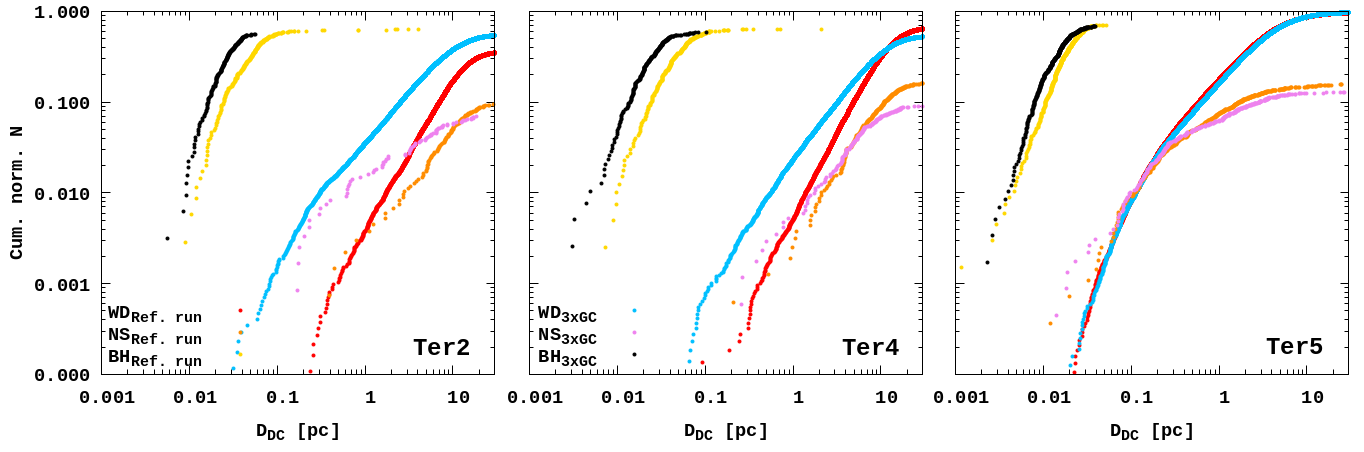}
\caption{Present-day distribution of high-mass stellar remnants, WDs, NSs, and BHs, for both types of runs. Red, orange, and yellow colours represent the {\tt ref. runs}. Light blue, violet, and black represent the {\tt common 3xGC run}.}
\label{fig:rem}
\end{figure*}

In Fig. \ref{fig:rem}, we show the present-day normalised cumulative number distributions of stellar remnants as a function of distance from the density centres of the GCs for the {\tt common 3xGC run} and the {\tt ref. runs}. In the case of the most massive GC, Ter5, we find almost no differences between these runs: the normalised cumulative distributions are nearly identical. We interpret this as a result of the very weak mutual gravitational influence of Ter2 and Ter4 on the internal structure of Ter5. In contrast, the central cumulative remnant distributions of Ter2 and Ter4 appear more compact in the {\tt common 3xGC run} than in the {\tt ref. runs}. This interesting behaviour of the high-mass remnant distribution can be understood in terms of the stronger gravitational influence of the massive Ter5 cluster on its lower-mass neighbours, Ter2 and Ter4. Due to periodic gravitational perturbations and stripping induced by Ter5, the internal distribution of remnants within the other Terzan clusters becomes slightly more concentrated, particularly for the WD population. 

%-------------------------------------------------------------------------%
%-------------------------------------------------------------------------%
%%%%%%%%%%%%%%%%%%%%%%%%%%%%%%%%%%%%%
\end{appendix}
%%%%%%%%%%%%%%%%%%%%%%%%%%%%%%%%%%%%%%%%

%%%%%%%%%%%%%%%%%%%%%%%%%%%%%%%%%%%%%%%%
\end{document}